\documentclass[showpacs,twocolumn,prl,amsmath,amssymb]{revtex4}

\usepackage{graphicx}
\usepackage{dcolumn}
\usepackage{bm}
\begin{document}
\title{
Opening of pseudogaps due to superconducting critical fluctuations\\
in quasi-two dimensions
}
\author{Fusayoshi J. Ohkawa}
\affiliation{Division of Physics, Graduate School of
Science,  Hokkaido University, Sapporo 060-0810, Japan}
\email{fohkawa@phys.sci.hokudai.ac.jp}
\received{~~ ~~ August 2005}
\begin{abstract} 
We examine the role of the anisotropy of  
superconducting thermal critical  fluctuations in the opening of pseudogaps  
in quasi-two dimensions. When 
the anisotropy of  coherence or correlation  lengths of the fluctuations 
is large enough and critical temperatures  $T_c$ are high enough, 
the energy dependence of the selfenergy
deviates from that of Landau's normal Fermi liquid 
in critical regions; its imaginary part has
no minimum at the chemical potential.
Prominent pseudogaps open in such a {\it non-Fermi liquid} phase.
Pseudogaps must be absent or subtle 
in quasi-two dimensional superconductors with low $T_c$
and almost isotropic three dimensional  superconductors. 
\end{abstract}
\pacs{74.20.-z, 74.90.+n, 71.10.-w}
\maketitle


Cuprate oxide superconductors, which are quasi-two dimensional ones,
exhibit high superconducting (SC) critical temperatures $T_c$ \cite{bednorz}.
On the other hand, no order is possible
at non-zero temperatures in one and two dimensions \cite{Mermin};
if $T_c$ were nonzero, 
thermal critical fluctuations would diverge at $T_c$.
Even if observed $T_c$ are high in quasi-low dimensional superconductors, 
they must be  $T_c$ substantially reduced by the fluctuations.
One may argue that some  anomalies must be accompanied by
the reduction of $T_c$.  It is plausible that the opening of
the so called spin-gap \cite{spingap} or the
pseudogaps \cite{Ding,Shen2,Shen3,Ino,Renner,Ido1,Ido2,Ekino}
is one of the accompanying anomalies. 
In actual, it has already been shown that the renormalization of quasiparticles
caused by SC fluctuations 
gives rise to the opening of pseudogaps \cite{pseudogap}.
The purpose of this paper is to clarify the role 
of the anisotropy of SC thermal critical fluctuations
in  the opening of pseudogaps in quasi-two dimensions.


In order to exhibit the essence of the issue in a simple manner,
we consider first  intermediate-coupling {\it attractive}
models on a quasi-two dimensional lattice
composed of square lattices:
\begin{equation}
{\cal H} = \sum_{ij\sigma} t_{ij} a_{i\sigma}^\dag a_{j\sigma}
+ \frac1{2} 
\sum_{ij \sigma\sigma^\prime} U_{ij}  
a_{i\sigma}^\dag a_{j\sigma^\prime} ^\dag 
a_{j\sigma^\prime} a_{i\sigma}  .
\end{equation}
When transfer integrals $t_{ij}$ between
nearest and next nearest neighbors on a plane,  $-t$ and $-t^\prime$,
are considered, the dispersion relation of electrons is given by
$E({\bf k}) = - 2 t \left[\cos(k_xa)+\cos(k_ya) \right]
- 4 t^\prime \cos(k_xa)\cos(k_ya) $,
with $a$ the lattice constant of square lattices;
the bandwidth is $8|t|$.
Denote attractive interactions $U_{ij}$ between onsite  and
nearest-neighbor pairs on a plane
by $U_0$ and $U_1$.
We consider two  models:
$U_0/|t | \simeq -4$ and $U_1 = 0$,  and 
$U_0 =0$ and $U_1/|t| \simeq -4$.
 Quasi-two dimensional features are phenomenologically considered 
as an anisotropy factor of SC fluctuations, which is introduced below.

When $U_0/|t|<0$ and $U_1=0$,
isotropic $s$-wave SC fluctuations are developed.
When $U_0=0$ and $U_1/|t|<0$, 
anisotropic $s$-wave, $p$-wave, or 
$d\gamma$-wave SC fluctuations can be developed.
We consider only two cases, isotropic $s$ wave ($\Gamma=s$)  and 
$d\gamma$ wave ($\Gamma = d\gamma)$.
When fluctuations of a single wave,
$s$ or $d\gamma$ wave, are considered,
the selfenergy to liner order in fluctuations is given by
 %
\begin{eqnarray}\label{EqSigma}
\Sigma_{\sigma}(i\varepsilon_n,{\bf k}) &=&
- \frac{k_BT}{N}\sum_{\omega_l{\bf q}}
U_{\Gamma}^2\eta_{\Gamma}^2
\left({\bf k}-\frac{1}{2}{\bf q}\right) 
\chi_{\Gamma}(i\omega_l,{\bf q}) 
\nonumber \\ && \times
G_{-\sigma}(-i\varepsilon_n-i\omega_l,-{\bf k}-{\bf q}) ,
\end{eqnarray}
%
where $N$ is the number of unit cells, 
$U_{\Gamma} =  U_0$ and 
$\eta_{\Gamma} ({\bf k}) =1$ for $\Gamma = s$,
$U_{\Gamma} =  U_1$ and 
$\eta_{\Gamma} ({\bf k}) =\cos(k_xa)-\cos(k_ya)$ 
for $\Gamma = d\gamma$, 
%
$G_{\sigma}(i\varepsilon_n, {\bf k}) =
1/[i\varepsilon_n + \mu - E({\bf k}) - 
\Sigma_{\sigma}(i\varepsilon_n,{\bf k}) ]$ 
is the renormalized Green function, 
with $\mu$ the chemical potential, and
$\chi_{\Gamma} (i\omega_l, {\bf q})$
is the SC susceptibility for $\Gamma$ wave.
%
%
%
%
The chemical potential shift caused by the Hartree and Fock terms 
is included in $\mu$.
Other types of fluctuations such as
charge ones also renormalize quasiparticles. 
They are ignored or a part of them
is phenomenologically considered as lifetime width $\gamma$,
which is introduced below.
Because critical fluctuations are restricted to a narrow region around
${\bf q}=0$, Eq.~(\ref{EqSigma}) is approximately given by
\begin{eqnarray}\label{EqSigma1}
\Sigma_{\sigma}(i\varepsilon_n,{\bf k}) &=&
- U_{\Gamma}^2\eta_{\Gamma}^2 \left({\bf k}\right)
k_B T \sum_{\omega_l}
G_{\sigma}^{(0)}(-i\varepsilon_n- i\omega_l, {\bf k}) 
\nonumber \\ && \times 
\ \frac1{N} \sum_{|{\bf q}_\parallel|\le q_c} \sum_{q_z}
\chi_{\Gamma}(i\omega_l,{\bf q}).
\end{eqnarray}
The summation over 
${\bf q}_\parallel=(q_x,q_y)$ is
restricted to $|{\bf q}_\parallel|\le q_c$, and
$G_{\sigma}^{(0)}(i\varepsilon_n, {\bf k}) =
1/\bigl[i\varepsilon_n +\mu -E({\bf k}) 
+ i \gamma \varepsilon_n/|\varepsilon_n|  \bigr]$
is used instead of 
$G_{\sigma}(i\varepsilon_n, {\bf k})$ to avoid a
selfconsistent procedure;
the lifetime width $\gamma$ is introduced
in $G_{\sigma}^{(0)}(i\varepsilon_n, {\bf k})$.
We also use a phenomenological expression for the SC susceptibility
to avoid another selfconsistent procedure:
\begin{equation}\label{EqPhChi}
\chi_{\Gamma}(i\omega_l,{\bf q}) =
\frac{\chi_{\Gamma}(0) \kappa^2}
{\displaystyle 
\kappa^2 + (q_\parallel a)^2 + \delta^2 ( q_z c)^2
+ \frac{|\omega_l|}{\Gamma_{SC}|t|} } ,
\end{equation}
with $\chi_{\Gamma}(0)$ the static homogeneous one,
$\delta$  the anisotropy factor of SC fluctuations, and
$c$  the  lattice constant along the $z$ axis. 
The so called $\omega$-linear term is ignored in Eq.~(\ref{EqPhChi}).
The density of states renormalized by SC fluctuations
is given by
\begin{equation}
\rho(\varepsilon) =
\frac1{N}  \sum_{\bf k}
\left(\! - \frac1{\pi} \right) 
\mbox{Im} \left[ G_{\sigma}^\prime(\varepsilon+i0,{\bf k}) \right] ,
\end{equation}
with
$G_{\sigma}^\prime(\varepsilon+i0,{\bf k})=
1/\bigl[1/G_\sigma^{(0)} (\varepsilon \!+\! i0,{\bf k}) 
- \Sigma_{\sigma}(\varepsilon \!+\! i0,{\bf k}) \bigr]$.


Following the previous paper \cite{pseudogap}, 
we can show that
$\chi_{\Gamma}(0) \kappa^2 |t| =O(1)$ when
$G_\sigma^{(0)} (\varepsilon \!+\! i0,{\bf k}) $ 
with $\gamma/|t| =O(1)$ is used 
in a microscopic theory.  We assume 
\begin{equation}
g_{\Gamma} \equiv
U_{\Gamma}^2 \chi_{\Gamma}(0) \kappa^2 /\pi |t| = 4 ,
\end{equation}
for the coupling constant $U_{\Gamma}$, $t^\prime = - 0.3 t<0$ 
for transfer integrals,
$\mu/|t|=-0.5$ for the chemical potential, 
$a=c$ for the lattice constants, 
$q_c=\pi/3a$ for the cut-off wave number, and
$\gamma/|t|=0.5$ for the lifetime width.

\begin{figure*}
\centerline{\hspace*{0.0cm}
\includegraphics[width=4.8cm]{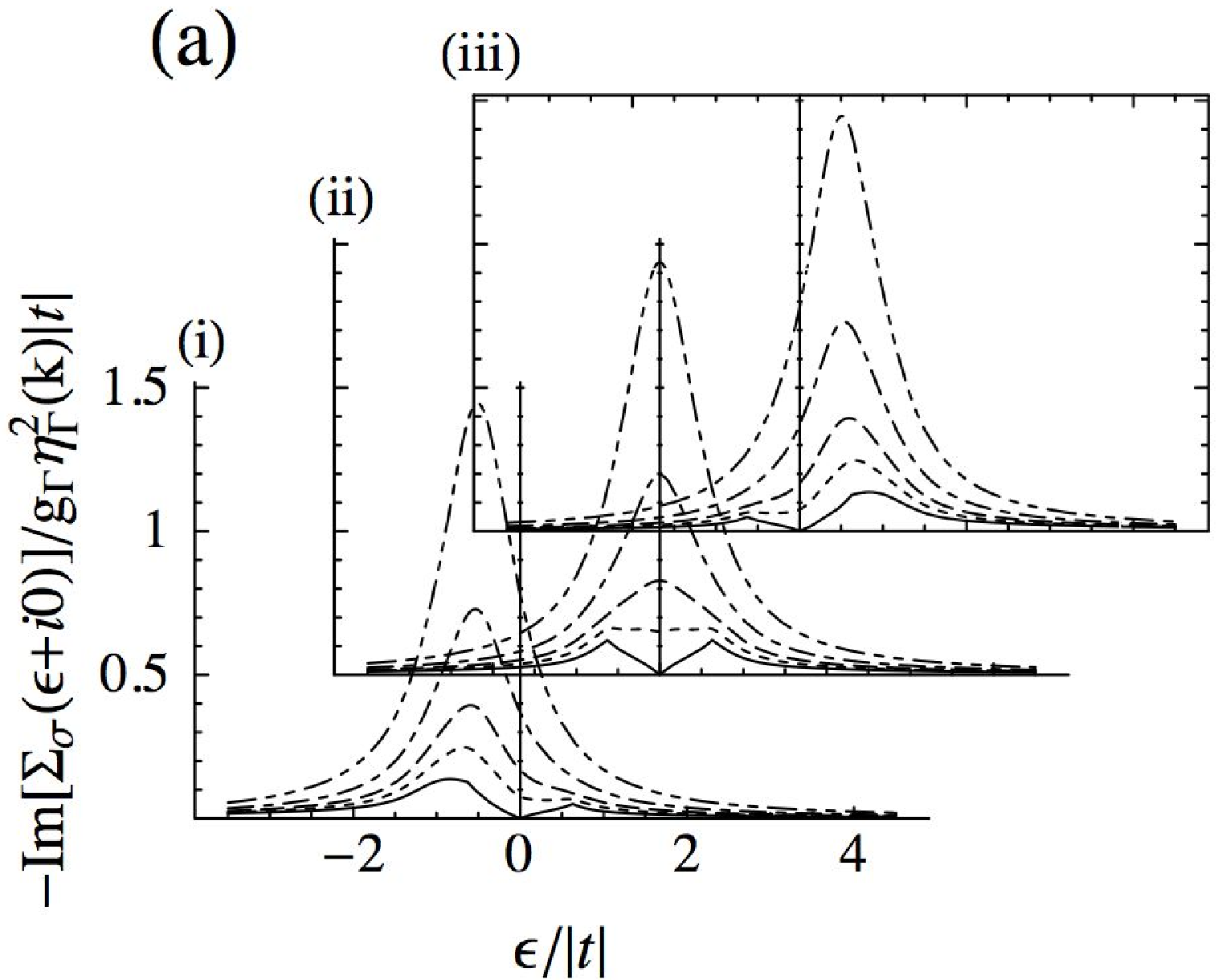}\hspace{-0.4cm}%
\includegraphics[width=4.8cm]{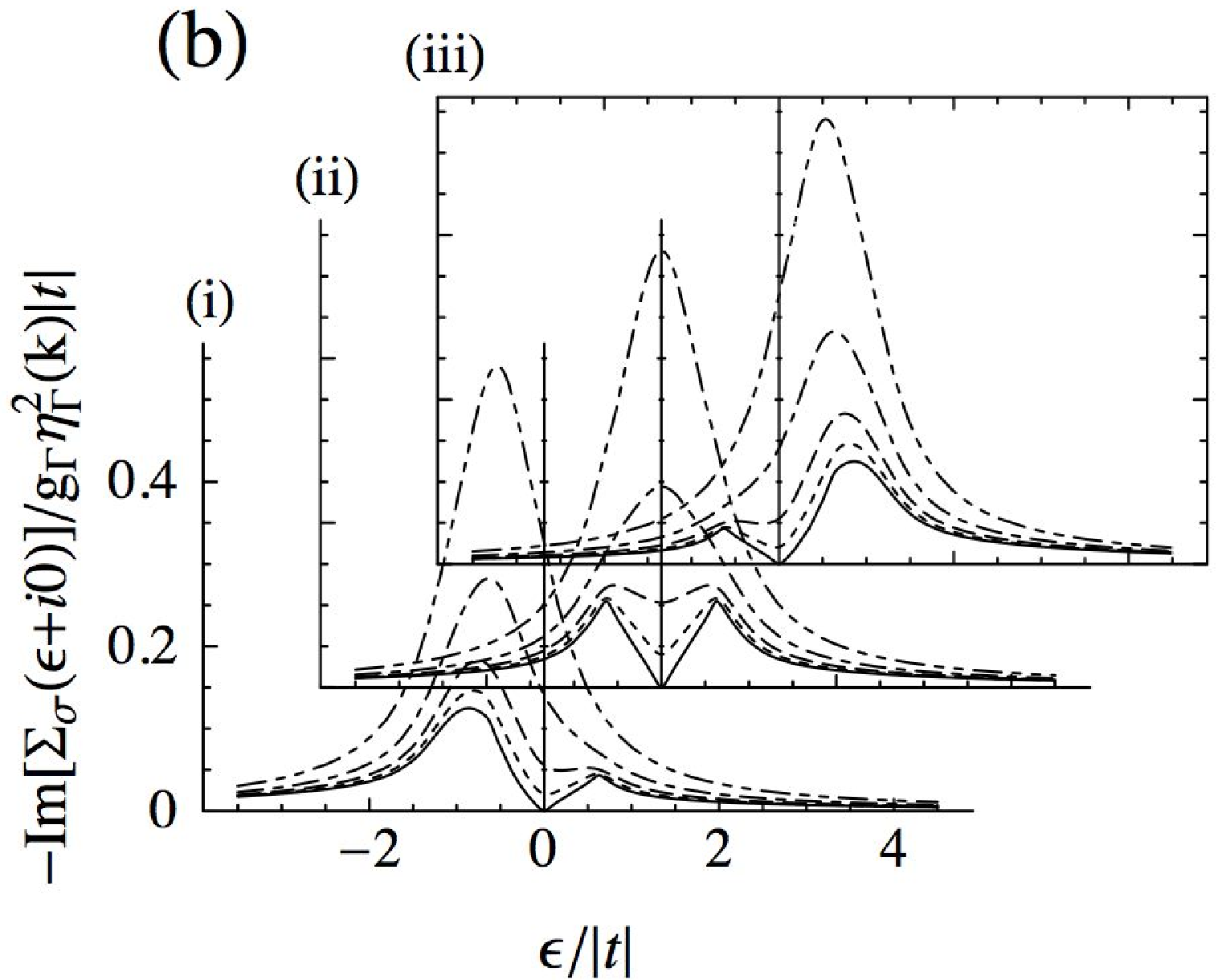}\hspace{-0.3cm}%
\includegraphics[width=4.8cm]{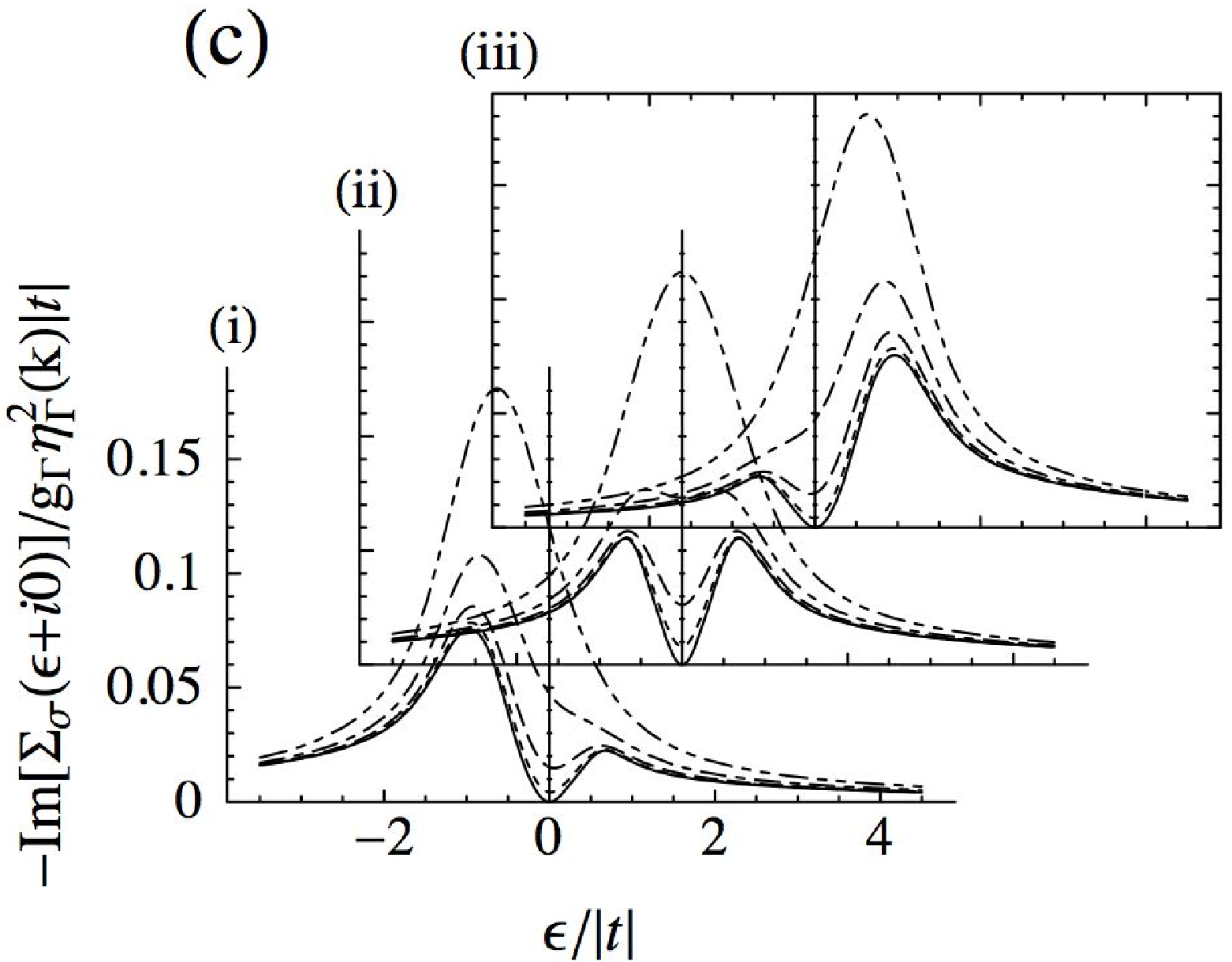}\hspace{-0.2cm}%
\includegraphics[width=4.8cm]{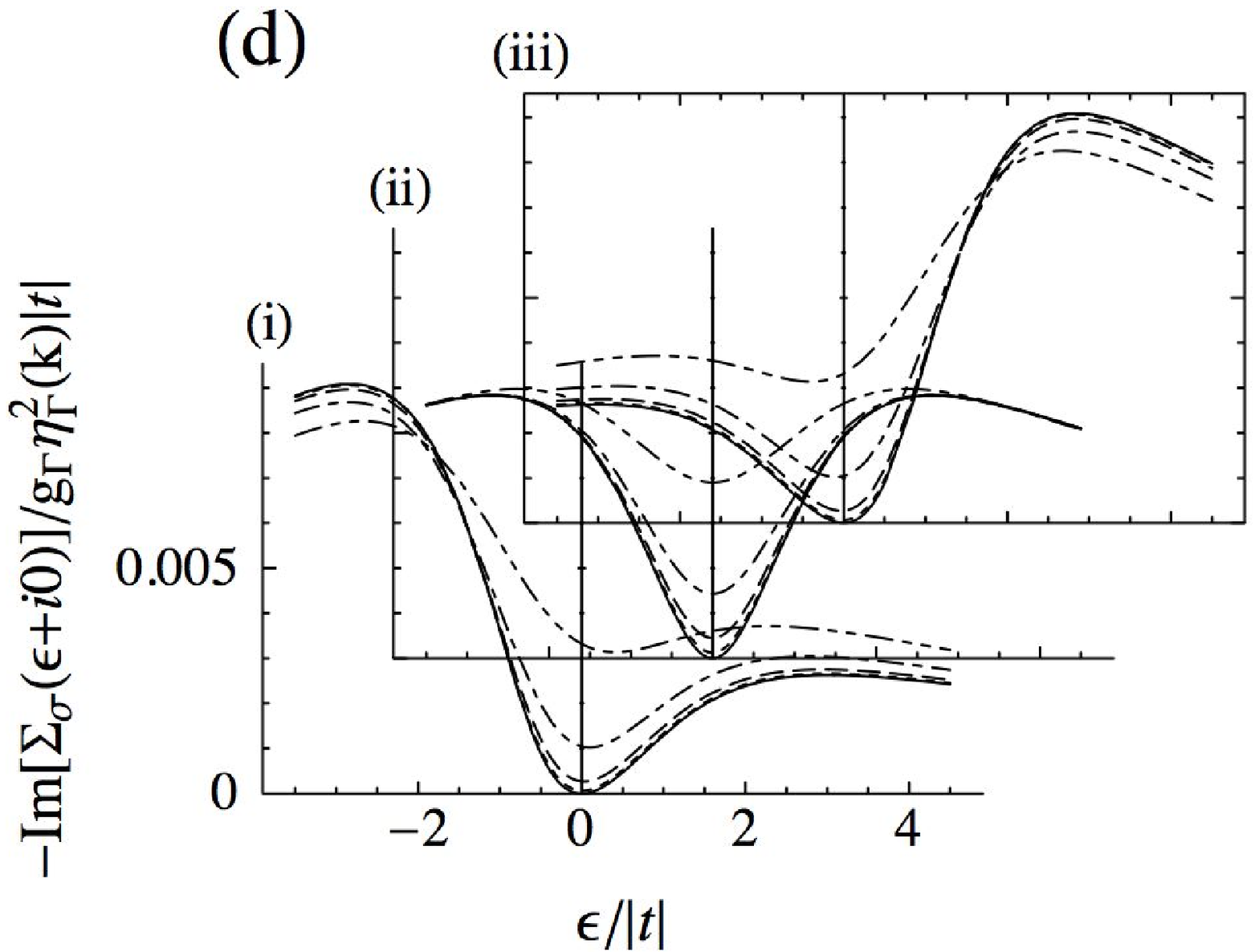}
}
\caption[2]{
$- \mbox{Im}\left[\Sigma_\sigma(\varepsilon+i0, {\bf k}) \right]/
g_{\Gamma}\eta_{\Gamma}^2({\bf k}) |t|$ 
at critical temperatures $(\kappa=0)$ for
$\Gamma_{SC}=0.2$ as a function of
$\varepsilon$: (a) $ \delta=0.01$, (b) $\delta=0.1$, (c) $\delta=0.3$,
and (d) $\delta=1$;
(i) $E({\bf k})=\mu -0.5|t|$, 
(ii) $E({\bf k})=\mu$, and 
(iii) $E({\bf k})=\mu +0.5|t|$.
In each figure, 
solid, dashed, broken, chain, and chain double-dashed
lines show results for $k_BT_c/|t|=0$,
0.05, 0.1, 0.2, 0.4, and 1, respectively.
The selfenergy correction is larger for smaller $\delta$.
When $\delta$ is small enough and $T$ is high enough,
the $\varepsilon$ dependence is different from that of conventional normal
Fermi liquids; there is no minimum at the zero energy or the chemical potential.
}
\label{Self}
\end{figure*}

First, we consider  SC critical points;
$T=T_c$ and $\kappa=0$.
Although $T_c$ depends on other parameters, we treat it  as
an independent one; our free parameters are 
$\delta$ and $\Gamma_{SC}$ in addition to $T_c$.
Figure~\ref{Self} shows
%
$- \mbox{Im}\left[\Sigma_\sigma(\varepsilon+i0, {\bf k}) \right]/
g_{\Gamma}\eta_{\Gamma}^2({\bf k}) |t|$
%
 as a function of $\varepsilon$ for three cases of $E({\bf k})-\mu$;
its ${\bf k}$ dependence comes through $E({\bf k})$. 
When fluctuations are isotropic $(\delta=1)$,
the selfenergy  is small and its $\varepsilon$
dependence is consistent with that of Landau's normal Fermi liquid,
as is shown in Fig.~\ref{Self}(d).
When fluctuations are anisotropic $(\delta \ll 1)$,
the selfenergy can be large, as are shown in Figs.~\ref{Self}(a)-(c).
Even when the anisotropy is large,  
the $\varepsilon$
dependence is consistent with that of the Fermi liquid
as long as $T$ is low,
as are shown in Figs.~\ref{Self}(a)-(c).
When the anisotropy is large and $T$ is high,
$\varepsilon=0$ is not any minimum point of 
$- \mbox{Im}\left[\Sigma_\sigma(\varepsilon \!+\! i0, E({\bf k}))\right]$,
as are shown in Figs.~\ref{Self}(a)-(c).
In such a {\it non-Fermi liquid} phase, quasiparticles on
the whole Fermi surface are incoherent in case of $s$ wave
and those around $(\pm\pi/a,0)$ and $(0,\pm\pi/a)$
are incoherent in case of $d\gamma$ wave.
Figure~\ref{Self} implies that 
the reduction of $T_c$ is large in quasi-two dimensional
superconductors if observed $T_c$ are high and
it is small in  almost isotropic three-dimensional ones
even if observed $T_c$ are high.

\begin{figure*}
\centerline{\hspace*{-0.0cm}
\includegraphics[width=6.0cm]{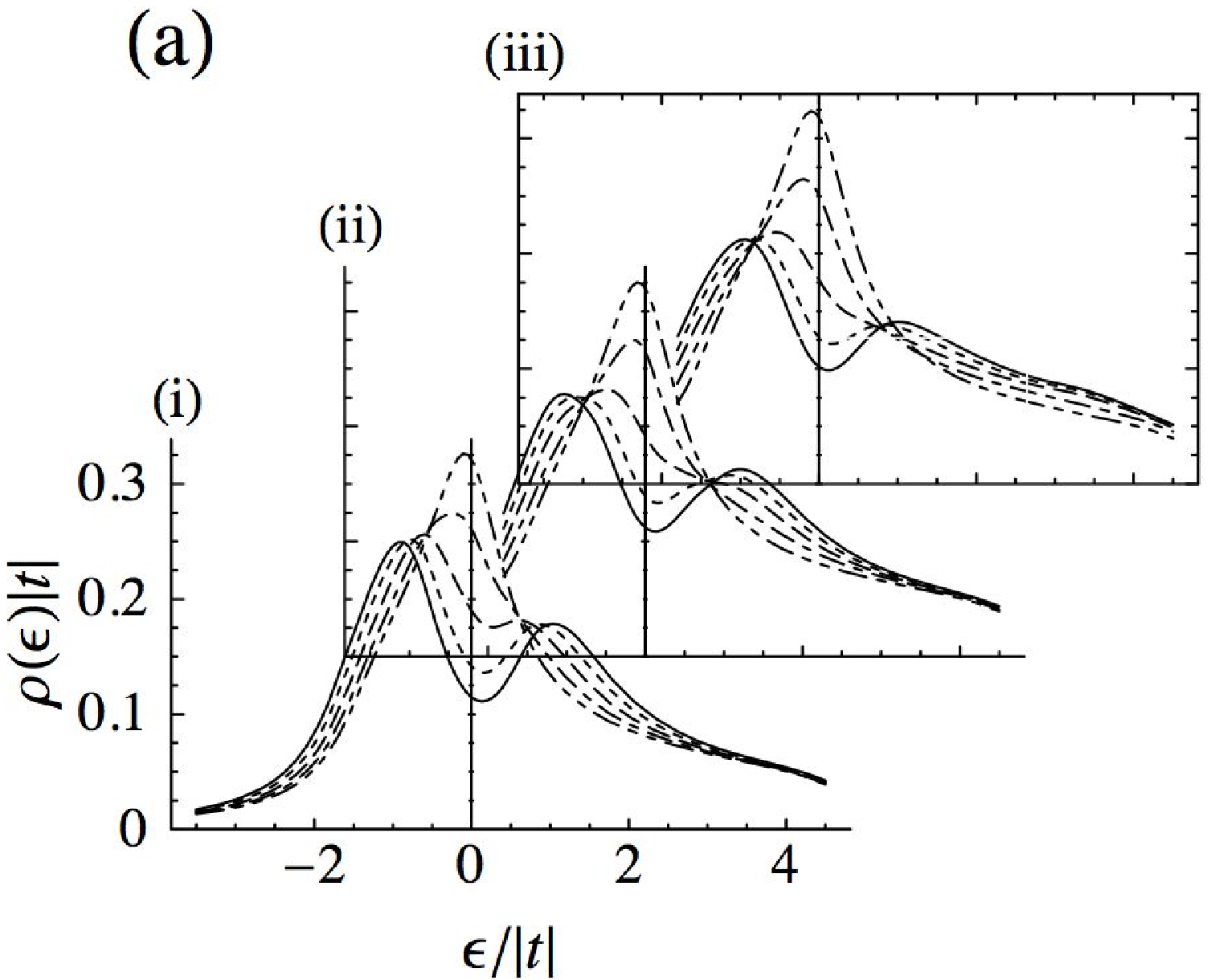}\hspace*{-0.3cm}
\includegraphics[width=6.0cm]{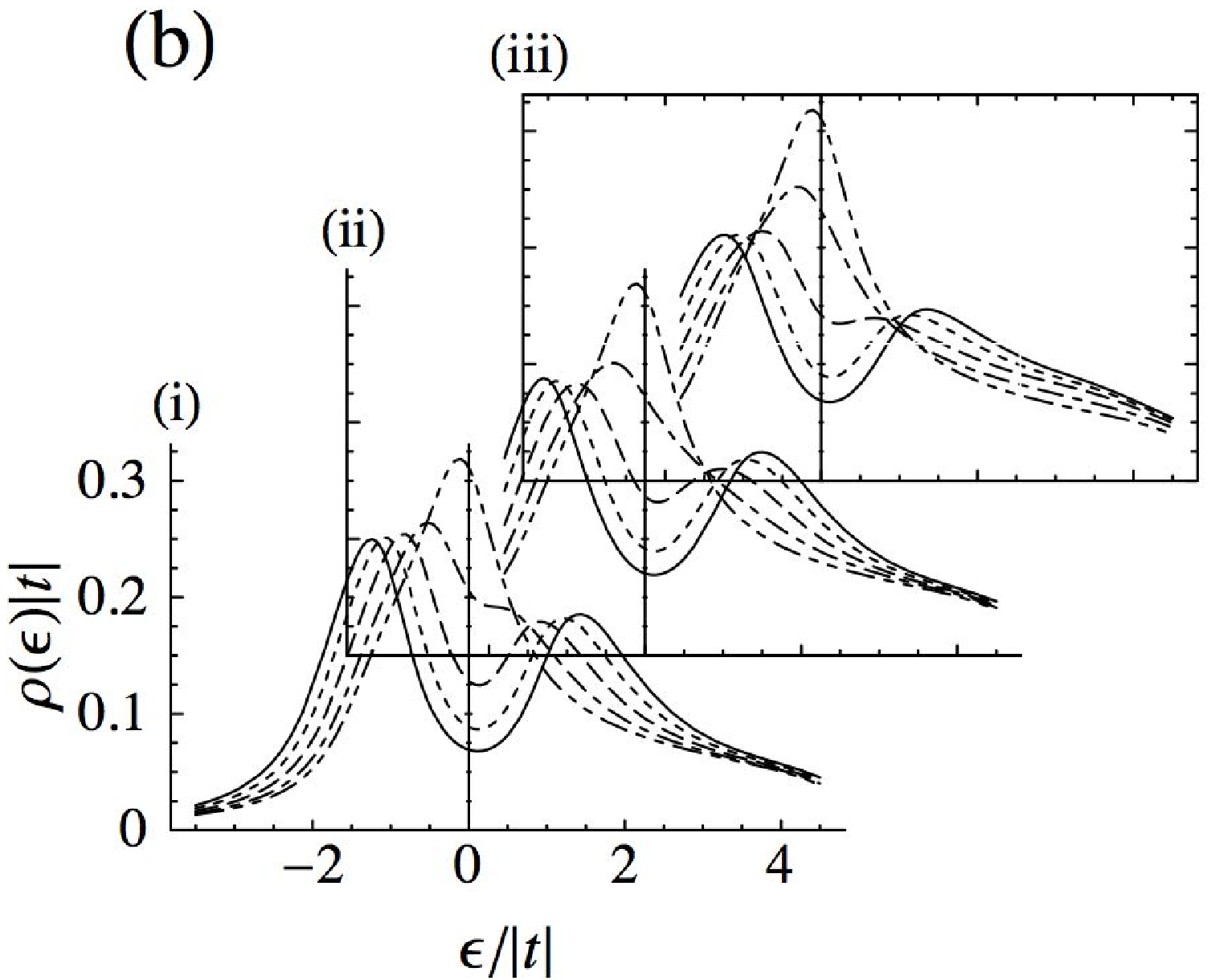}\hspace*{-0.3cm}
\includegraphics[width=6.0cm]{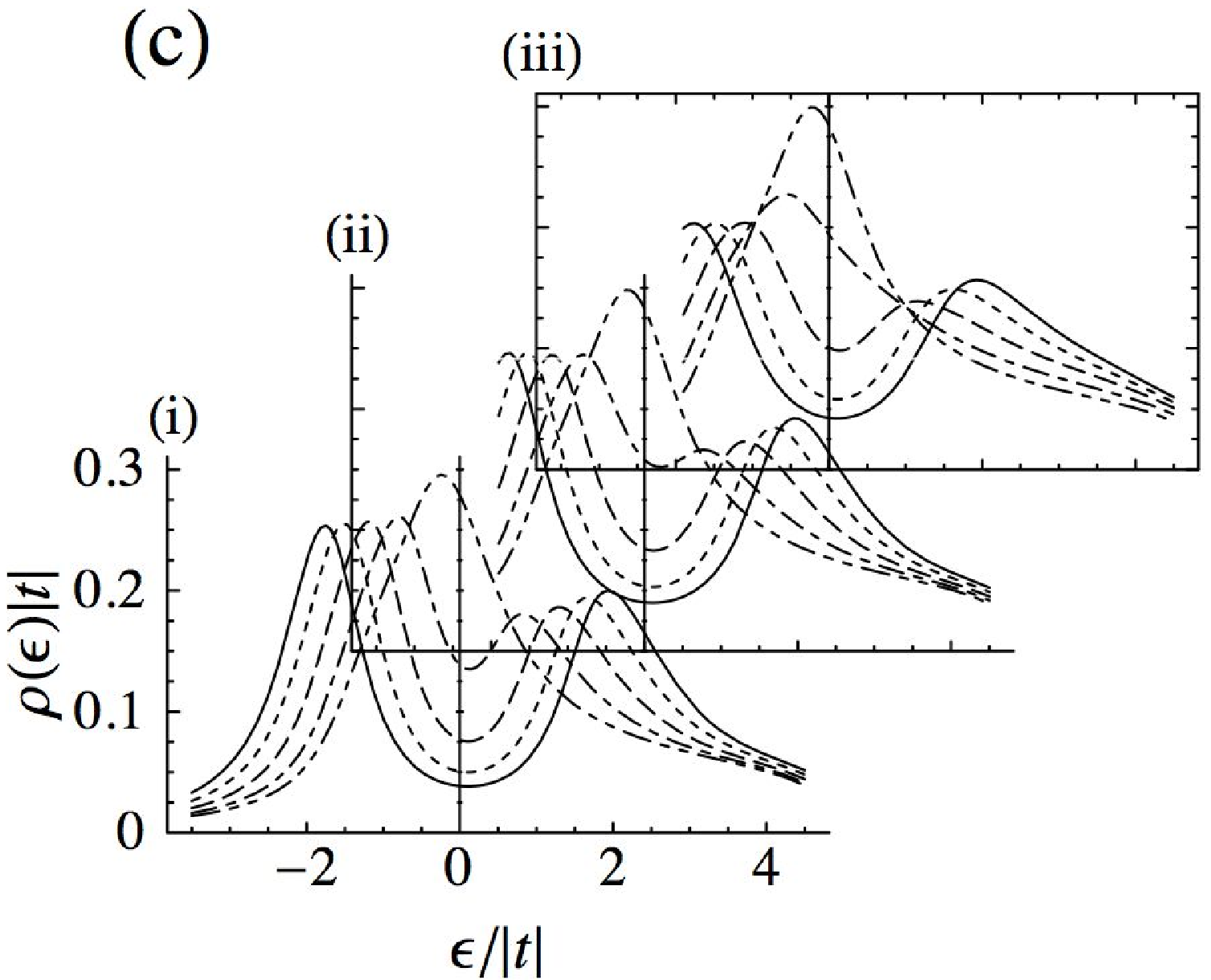}
}
\caption[2]{
$\rho(\varepsilon)$
at critical temperatures $(\kappa=0)$
for $s$-wave and 
$g_{s}=4$: (a) $k_BT_c/|t|=0.1$, (b) $k_BT_c/|t|=0.2$,
and (c) $k_BT_c/|t|=0.4$; (i) $\Gamma_{SC}=0.1$, (ii) $\Gamma_{SC}=0.3$,
and (iii) $\Gamma_{SC}=1$.
In each figure, solid, dashed, broken, chain, 
and chain double-dashed lines show
$\rho(\varepsilon)$ for $\delta=0.01$, 0.03, 0.1, 0.3, and 1,
respectively. Pseudogaps are more prominent for higher  $T_c$, smaller $\delta$,
and smaller $\Gamma_{SC}$.
Pseudogaps are absent in any spectrum for the isotropic case $(\delta=1)$.
}
\label{rho-s}
\end{figure*}

\begin{figure*}
\centerline{\hspace*{0.0cm}
\includegraphics[width=6.0cm]{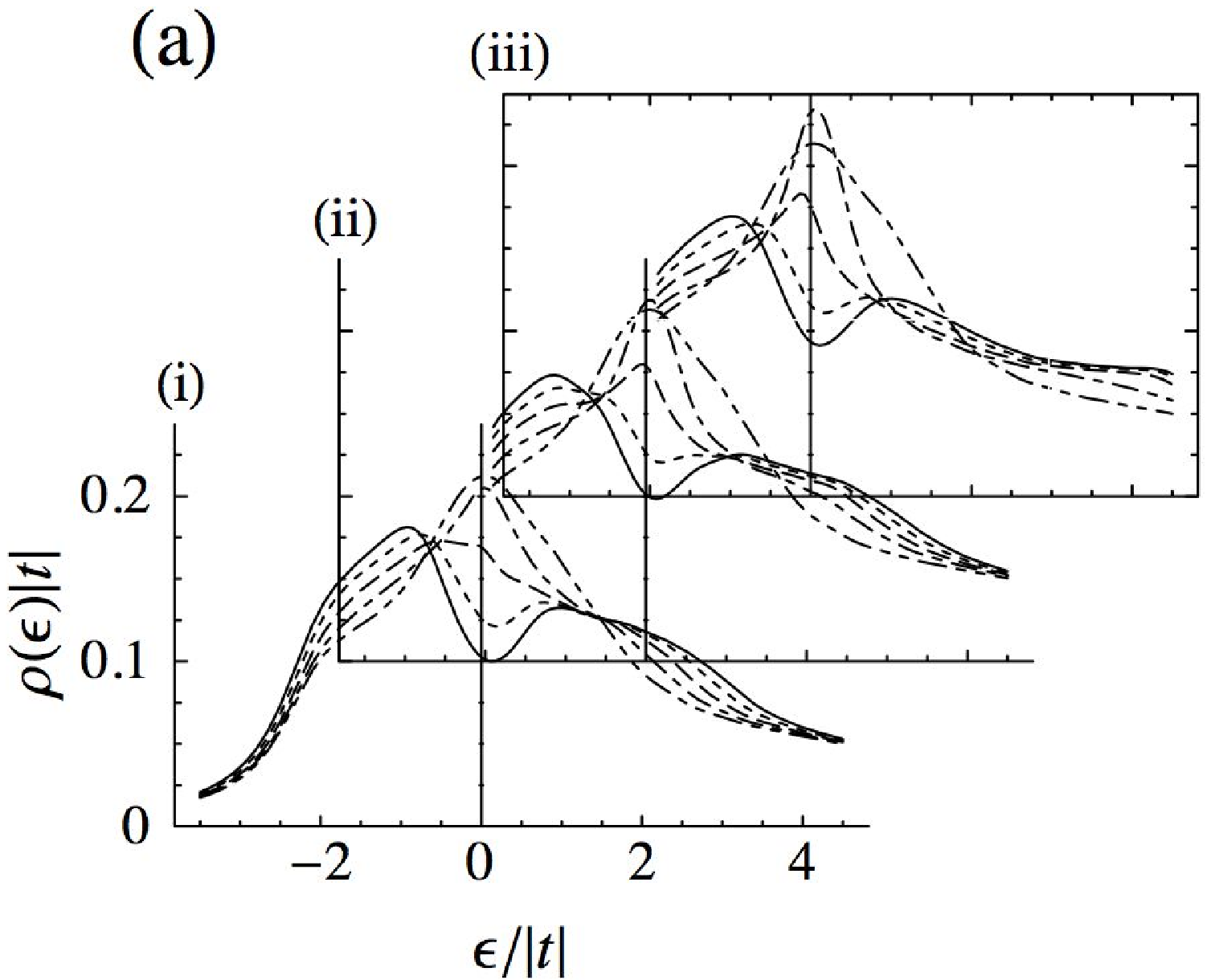}\hspace{-0.3cm}%
\includegraphics[width=6.0cm]{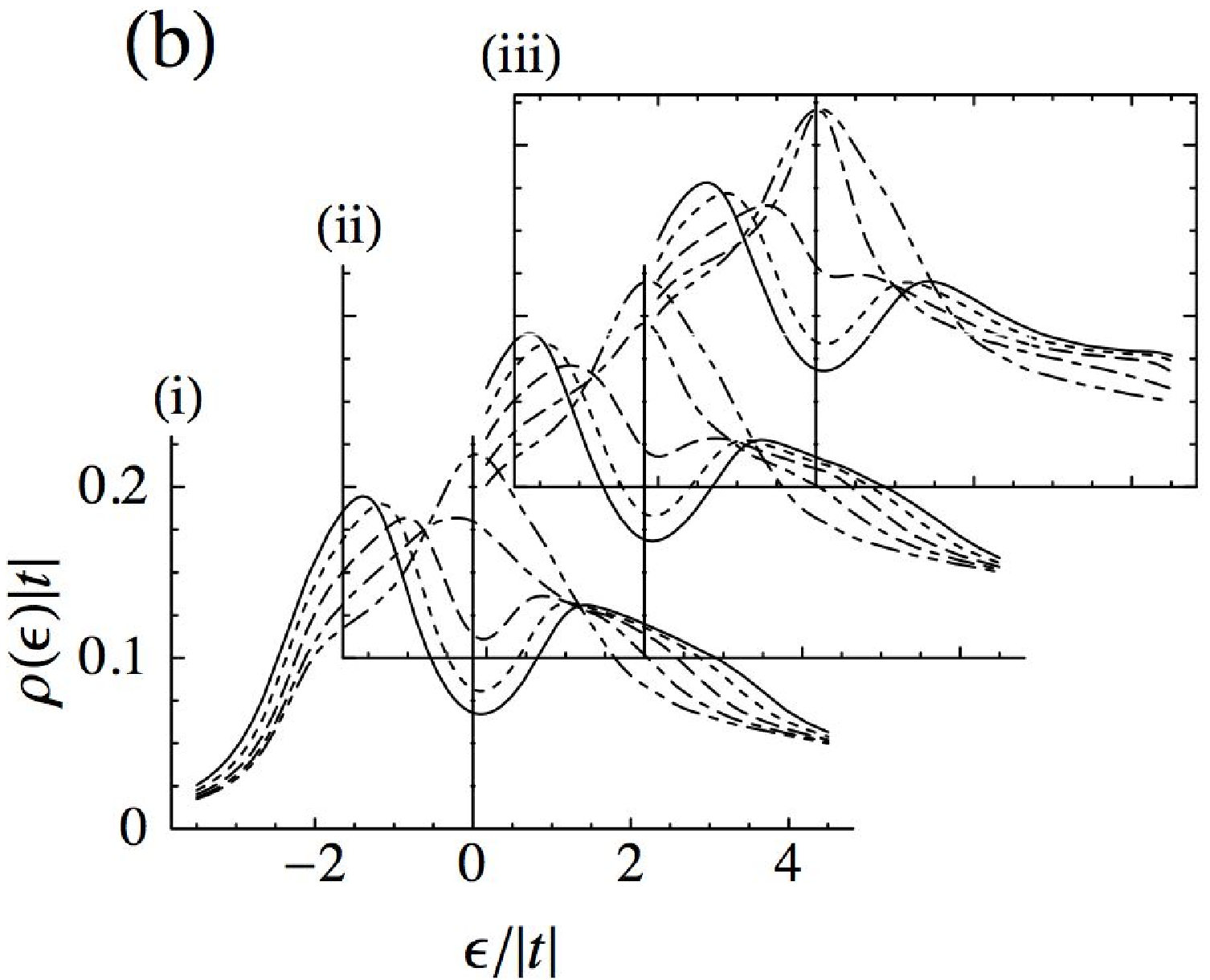}\hspace{-0.3cm}%
\includegraphics[width=6.0cm]{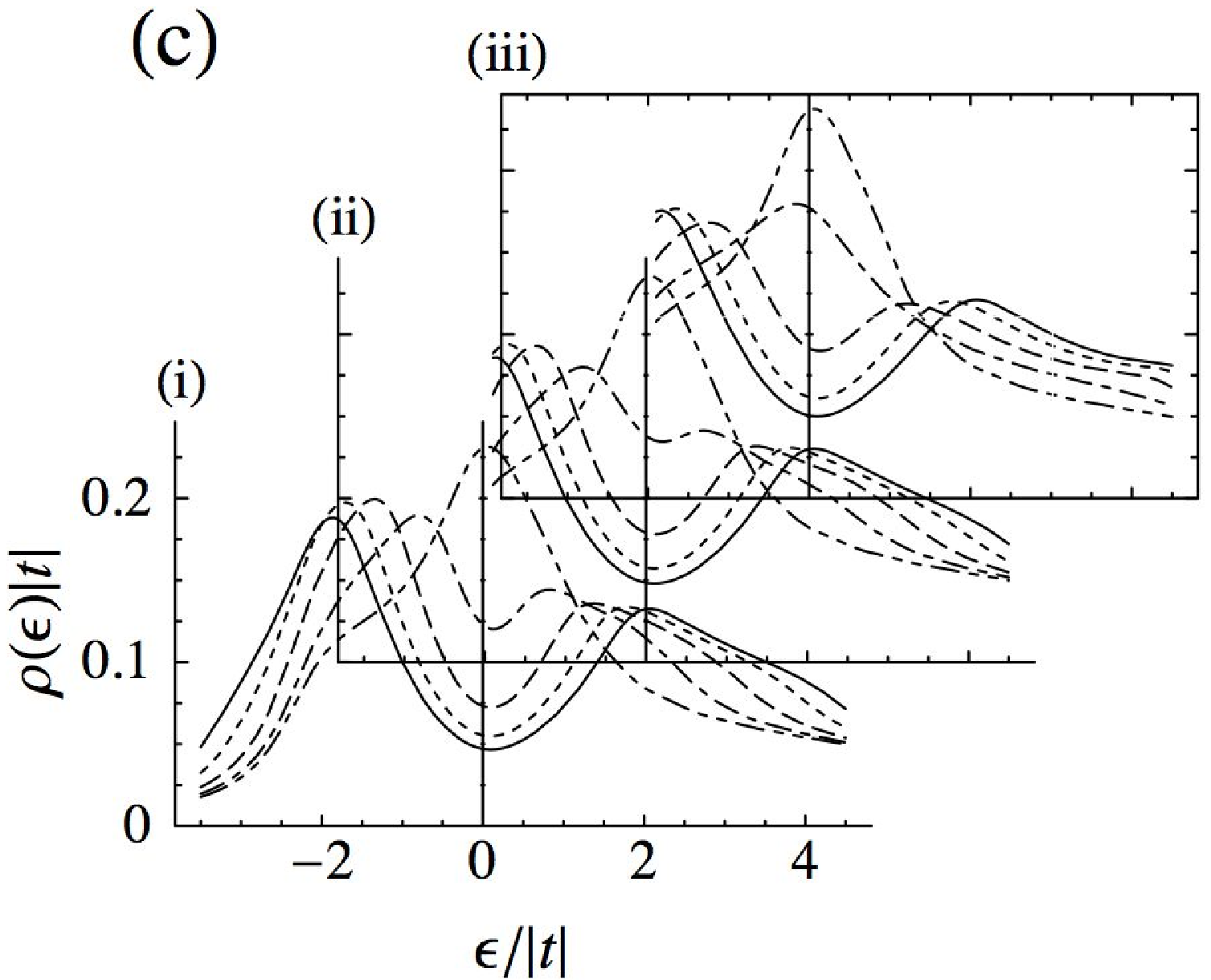}
}
\caption[3]{
$\rho(\varepsilon)$
at critical temperatures  $(\kappa=0)$ for $d$-wave and $g_{d\gamma}=4$.
See  also the figure caption of Fig.~\ref{rho-s}. 
No essential difference can be seen between Fig.~\ref{rho-s} for $s$ wave
and this figure for $d\gamma$ wave.
}
\label{rho-d}
\end{figure*}
\begin{figure*}
\centerline{\hspace*{0.0cm}
\includegraphics[width=5.0cm]{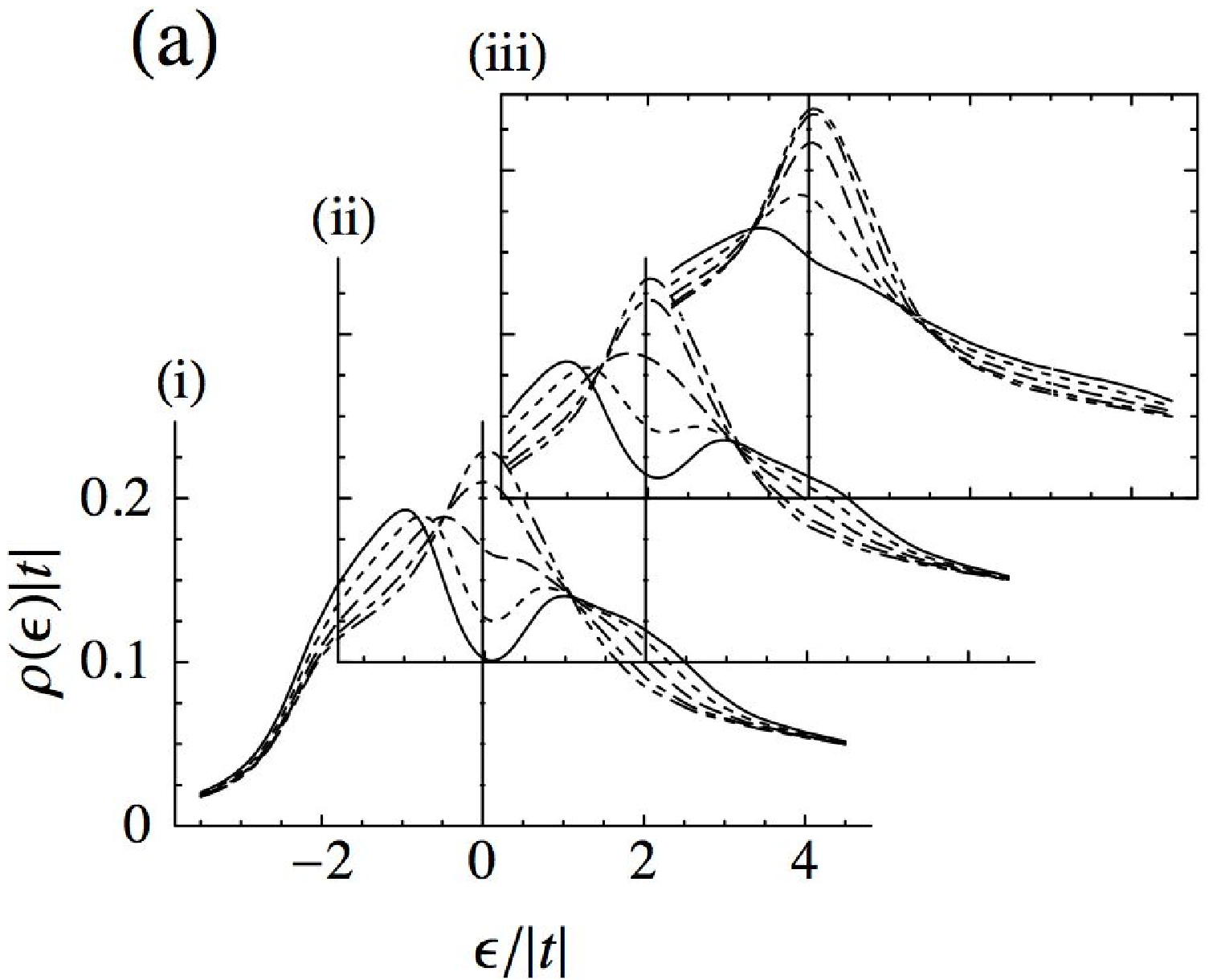}\hspace{-0.6cm}%
\includegraphics[width=5.0cm]{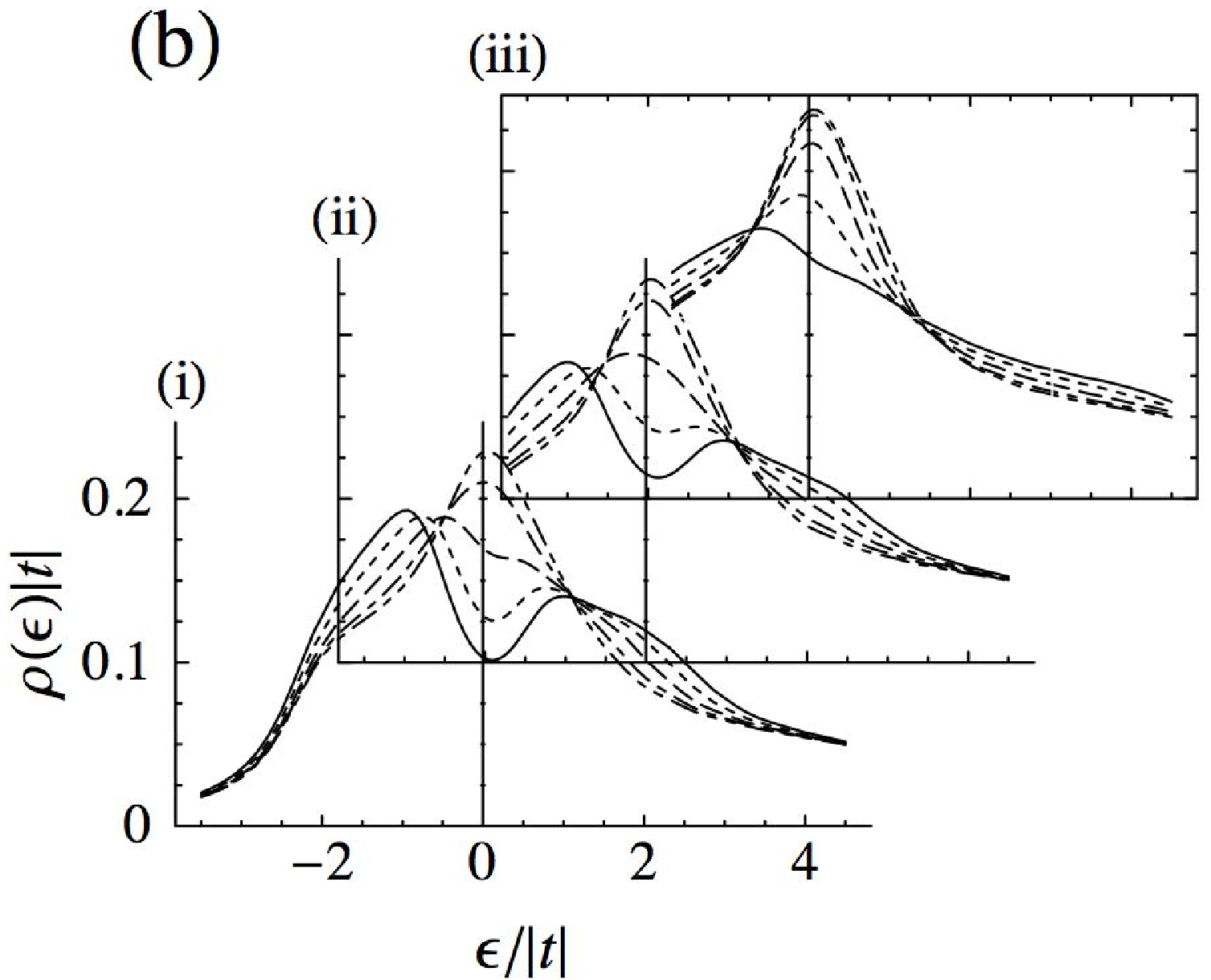} \hspace{-0.6cm}%
\includegraphics[width=5.0cm]{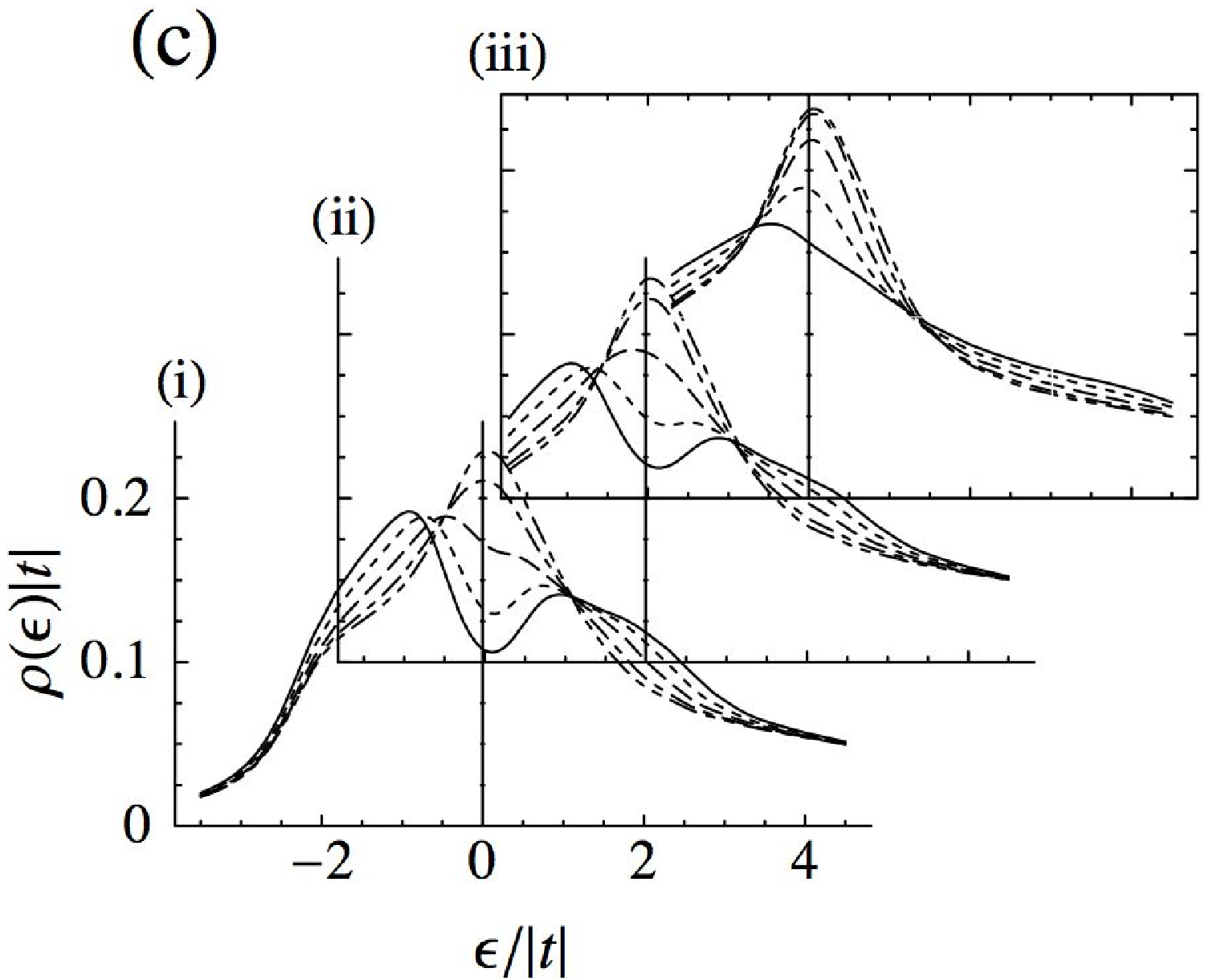}\hspace{-0.6cm}%
\includegraphics[width=5.0cm]{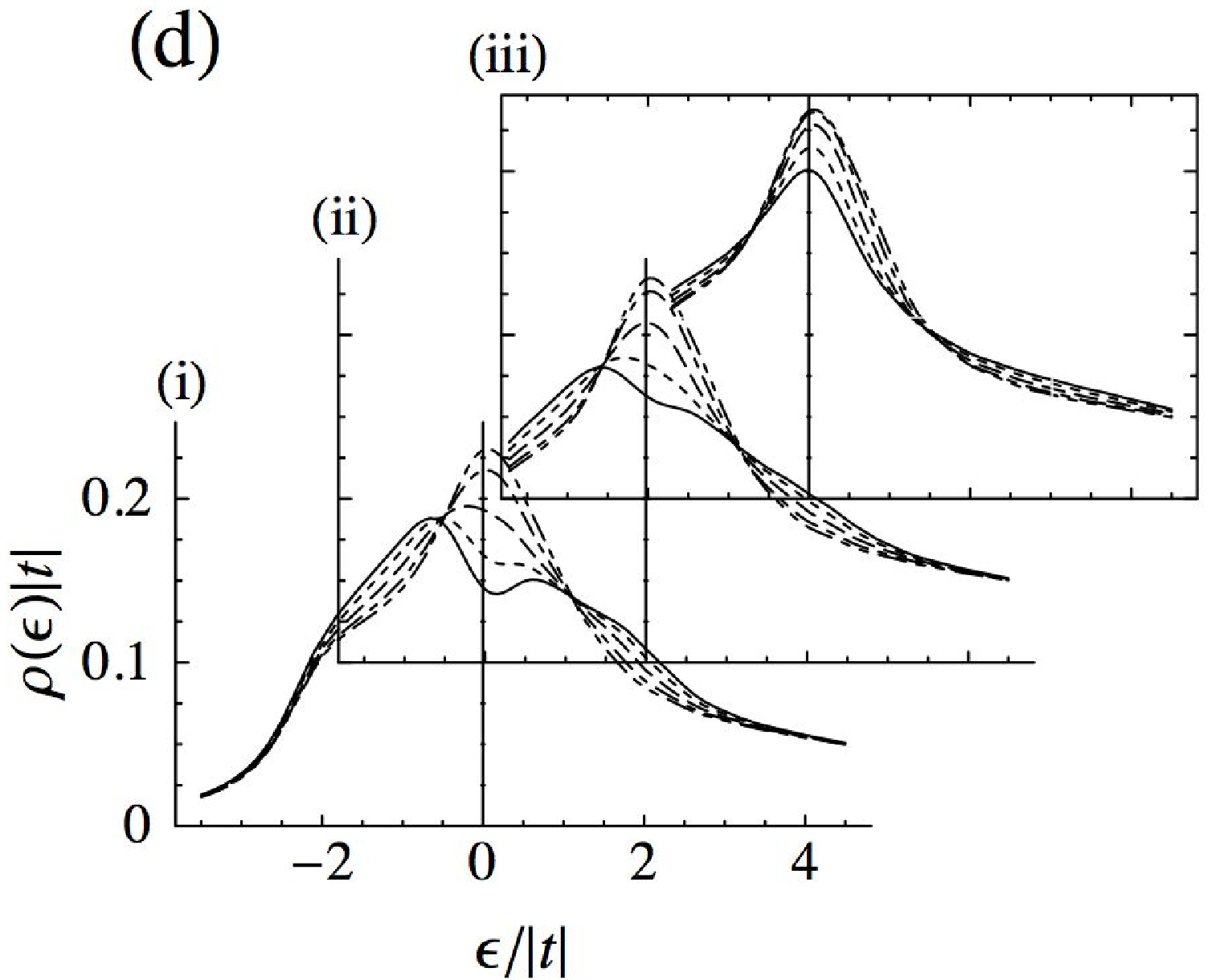}
}
\caption[4]{
$\rho(\varepsilon)$
at $k_BT/|t|=0.4$ for $d\gamma$ wave and $g_{d\gamma}=4$:
(a) $\delta=0.01$, (b) $\delta=0.03$, (c) $\delta=0.1$, and (d) $\delta=0.3$;
(i) $\Gamma_{SC}=0.1$, (ii) $\Gamma_{SC}=0.3$, and
(iii) $\Gamma_{SC}=1$.
In each figure, 
solid, dashed, broken, chain, and chain double-dashed lines
show $\rho(\varepsilon)$ for $\kappa^2=0.5$,
1, 2, 4, and 8, respectively. 
When $\kappa^2$ or/and $\Gamma_{SC}$ are large enough,
pseudogaps  are absent or subtle. 
For example, pseudogaps are absent or subtle in any spectrum for $\kappa^2 \ge 2$;
they are absent  in any spectrum for $\Gamma_{SC}=1$.
}
\label{rho>Tc}
\end{figure*}

Figures~\ref{rho-s} and \ref{rho-d} show $\rho(\varepsilon)$
renormalized by $s$-wave and $d$-wave
SC fluctuations, respectively.
Prominent pseudogaps open in the {\it non-Fermi liquid} phase.
Pseudogaps  are  more prominent for higher  $T_c$.
Because higher $T_c$ are mainly caused by larger $g_\Gamma$,
the tendency that pseudogaps  are more prominent for higher $T_c$
must be  larger in actual superconductors than it is 
in Figs.~\ref{rho-s} and \ref{rho-d}. Large anisotropy of
critical fluctuations or a small $\delta$ such as 
$\delta < 0.1$-0.3 is  indispensable for the opening of
prominent pseudogaps;
pseudogaps  are absent or  subtle in the isotropic case $(\delta=1)$.
 Although spectra of $\rho(\varepsilon)$ are slightly 
different between $s$-wave and $d$-wave cases, there is no
essential difference between them as long as $U_{s} \simeq U_{d\gamma}$.
The anisotropy of
critical fluctuations within planes plays a minor role.


When cuprate oxides  are
treated, the theory should be extended to 
the strong-coupling {\it repulsive} regime, 
$U_{0}/|t| \agt 8$  and $|U_1/t|\ll 1$; we had better use
the so called $d$-$p$ or the $t$-$J$ model \cite{ZhangRice}.
A three-peak structure of
the so called Gutzwiller's quasiparticle band \cite{Gutzwiller}  between
the lower and upper Hubbard bands (LHB and UHB) \cite{Hubbard}
is mainly because of strong local correlations \cite{slave}.
An intersite exchange interaction  can be 
a Cooper-pair interaction \cite{Hirsch}.
The main part of the exchange interaction 
in cuprate oxides is the superexchange interaction, 
which arises from the virtual exchange
of pair excitations of electrons across LHB and UHB.
An exchange interaction arising from that 
of Gutzwiller's quasiparticles also
work between Gutzwiller's quasiparticles themselves.
The strong local correlations, which give rise to the three-peak structure,
 and the exchange interactions
can be treated by a Kondo-lattice theory 
\cite{pseudogap,KL-theory1,KL-theory2}.
According to 
the single-site approximation (SSA) \cite{Mapping-1,Mapping-2}
or the dynamical mean-field theory (DMFT) \cite{georges},
the three-peak structure is mapped to the so called Kondo peak
between two subpeaks in the Anderson model
or the Kondo problem.  In the  Kondo-lattice theory,
an {\it unperturbed} state is
constructed by including the local correlations in SSA or DMFT and intersite effects
are perturbatively considered starting from the unperturbed state.
Because SSA or DMFT is rigorous for infinite dimensions  \cite{metzner},
this perturbation is nothing but $1/d$ expansion,
with $d$ spatial dimensionality.
 Then,  a theory of superconductivity or SC fluctuations
in the vicinity of the Mott transition or crossover
can be developed almost in parallel to that of this paper
\cite{pseudogap,KL-theory1,KL-theory2}.
Eventually, 
$t$ of this paper is replaced by that of Gutzwiller's quasiparticles, and 
the attractive interaction of this paper
is replaced by the exchange interaction between
nearest neighbors in such a way that
%
$U_{d\gamma} \rightarrow U_{d\gamma}^*=\frac{3}{4}I_s \bigl(
\tilde{\phi}_s / \tilde{\phi}_\gamma \bigr)^2$,
%
with $\tilde{\phi}_s$ and $\tilde{\phi}_\gamma$ 
the single-site vertex correction in spin channels and
the mass renormalization factor in SSA or DMFT.
Here, $I_s$ is the exchange interaction constant between nearest neighbors,
whose main part is $J_s$ of the superexchange interaction.
Phenomenological $\gamma$ is mainly due to antiferromagnetic spin fluctuations
instead of charge ones.
Because $T_c$ of  $d\gamma$ wave are much higher than $T_c$
of other waves, we consider only
$d\gamma$ wave in the following part. 

The specific heat coefficient  of the so called optimal doped
cuprate-oxide superconductors is as large as 
$14$~mJ/K$^2$mol \cite{gamma1}. This implies that
the Gutzwiller's quasiparticle band width is about 0.3~eV or $|t|\simeq 0.04$~eV.
On the other hand, the superexchange interaction 
is as strong as $|J_s| \simeq 0.1$-0.15~eV.
Therefore, high-$T_c$ superconductivity 
must occur in an {\it intermediate} coupling regime
$|U_{d\gamma}^*/t| \simeq 4$.
Because $T_c/|t|=0.2$ corresponds to $T_c\simeq 100$~K 
and  $\delta \alt 0.1$ in cuprate oxides,
Fig.~\ref{rho-d}(b) implies that
the opening of pseudogaps at $T_c$ must be caused
by SC thermal critical fluctuations;
even if other mechanisms work, the mechanism proposed
here must play a major role, at least, at $T_c$ and in critical regions.

If all the parameters such as $\chi_{\Gamma}(0)$,
$\kappa$, $\delta$, and $\Gamma_{SC}$,
 were constant as a function of $T$,
pseudogaps would be developed with increasing $T$.
Experimentally, however, pseudogaps close at high enough $T$.
It is likely that the temperature dependences of 
$\chi_{\Gamma}(0)$,
$\kappa$, $\delta$, and $\Gamma_{SC}$ are responsible for
the closing of  pseudogaps, for example, 
at $k_BT/|t|\simeq 0.4$ or $T\simeq 200$~K.
Then, we examine what parameters are needed
in order to reproduce a situation that
pseudogaps that open at $k_BT/|t|=0.2$ close at $k_BT/|t|=0.4$.
It is obvious that $\kappa^2$ increase with
increasing $T$; $\chi_{\Gamma}(0)\kappa^2$ is almost
constant. It is likely that $\Gamma_{SC}$
increase with increasing $T$.
Figure~\ref{rho>Tc} show that
when $\kappa^2$ and/or $\Gamma_{SC}$ are large enough
at $k_BT/|t|=0.4$ no prominent pseudogap can be seen.
For example, no prominent pseudogap can be seen
for $\kappa^2 \agt 2$ even when $\Gamma_{SC}=0.1$.
It is interesting to complete the
selfconsistent procedure, which is avoided in this paper,
in order to confirm whether or not such temperature dependences
of $a/\kappa$ and $\Gamma_{SC}|t|$,  
the correlation length and the energy scale of fluctuations,
 can be actually reproduced.

The so called coherence peaks are missing
in Figs.~\ref{rho-s}, \ref{rho-d}, and \ref{rho>Tc}. 
We also note 
that  $\rho(0)|8t| = O(1)$ even when pseudogaps are developed.
This is quite different from $\rho(0)|8t| \ll 1$
 at $T\ll T_c$ in SC phases.
The origin of the so called zero-temperature pseudogap
\cite{Davis},
which is characterized by $\rho(0)|8t| \ll 1$, 
must be different from that proposed in this paper.

The density of states in SC phases of cuprate-oxide
superconductors is different from that 
of conventional ones:
dips outside SC gaps in cuprate oxides 
and no dips in conventional ones \cite{Renner,Ido1,Ido2}.
Low-energy SC fluctuations $|\omega| \alt \varepsilon_G(T)$, with
$\varepsilon_G(T)$ being SC gaps, must be suppressed below $T_c$,
but high-energy fluctuations such as
$|\omega| \simeq \varepsilon_G(T)$ or
$|\omega| \agt \varepsilon_G(T)$ must be developed. 
Pseudogaps  may  open at high-energy regions
because of  high-energy SC fluctuations
even  in SC phases.
A possible scenario is that dips appear 
because of the superposition of 
SC gaps  and pseudogaps.

Because low-energy SC fluctuations 
$|\omega| \alt \varepsilon_G(0)$
cannot developed at $T=0$~K,
the reduction of $\varepsilon_G(0)$
at $T=0$~K must be very small.
In the mean-field approximation for $d\gamma$ wave
\cite{KL-theory1}, where
the reduction of $T_c$ is not considered, 
$\varepsilon_G(0)/k_BT_c \simeq 4.35$.   
Observed  large ratios \cite{Renner,Ido1,Ido2}
 of  $\varepsilon_G(0)/k_BT_c \agt 8$
are pieces of evidence that $T_c$ are actually reduced by 
SC critical fluctuations. 
  %

Transition-metal dichalcogenide and organic 
superconductors  are another possible low dimensional 
high-$T_c$ ones \cite{layer2,layer3}. If $T_c$ are high and 
$\varepsilon_G(0)/k_BT_c$ are large, pseudogaps
must also be prominent in critical regions. 

It is straightforward to extend
the theory of this paper to pseudogaps due to 
spin and charge fluctuations.
When $T_c$ of spin or charge density wave 
are  high and the anisotropy of fluctuations is large,
pseudogaps must also be prominent in critical regions.


Mercury-based cuprate oxides show very high $T_c$ under pressures
\cite{ott,chu}.
Pressures must reduce the anisotropy so that
the reduction of $T_c$ becomes smaller with increasing pressures.
It is interesting to search for almost isotropic cuprate-oxide 
superconductors  with no prominent pseudogap.  
Because the reduction of $T_c$ by critical fluctuations is small,
their $T_c$ can be higher than $T_c$ of quasi-two dimensional ones.
A simple argument implies that
if $\varepsilon_G(0)/k_BT_c = $4-5 are realized
$T_c$ can exceed 200~K.


In conclusion,
pseudogaps can open in critical regions of
quasi-two dimensional superconductors
with high critical temperatures $T_c$ because of 
thermal critical fluctuations, which can be well developed
only in low dimensions. 
They can open for 
not only anisotropic superconductors but also isotropic or BCS ones
in quasi-two dimensions.
Because it is difficult for the fluctuations to be developed 
in quasi-two dimensional superconductors with low $T_c$ and
almost isotropic three dimensional  ones,  even with high $T_c$,
it is unlikely that prominent  pseudogaps open in such 
superconductors.





\begin{thebibliography}{}
\bibitem{bednorz}
J.G. Bednortz and K.A. M\"{u}ller, 
Z. Phys. B {\bf 64}, 189 (1986).
%
\bibitem{Mermin}
N.D. Mermin and H. Wagner,
Phys. Rev. Lett. {\bf 17}, 1133 (1966).
%
\bibitem{spingap}
H. Yasuoka, et al.,
{\it Strong Correlation and Superconductivity},
Springer Series in Solid State Science Vol. 89
(Springer-Verlag, Berlin, New York 1989), p. 254.
%
\bibitem{Ding}
H. Ding, et al.,
Nature {\bf 382}, 51 (1996).  
%
%
\bibitem{Shen2} 
J. M. Harris, et al.,
Phys. Rev. B {\bf 54}, R15665 (1996).   
%
\bibitem{Shen3} 
A. G. Loeser,  et al.,
Science {\bf 273}, 325 (1996).  
%
\bibitem{Ino}
A. Ino, et al.,
Phys. Rev. B {\bf 65}, 094504 (2002).
%
\bibitem{Renner}
Ch. Renner, et al.,
Phys. Rev. Lett. {\bf 80}, 149 (1998). 
%
\bibitem{Ido1}
M. Oda, et al.,
Physica C {\bf 281}, 135 (1997).
%
\bibitem{Ido2}
T. Nakano, et al.,
J. Phys. Soc. Jpn. {\bf 67}, 2622 (1998).
%
%
%
\bibitem{Ekino}
T. Ekino, et al., 
Phys. Rev. B {\bf 60}, 6916 (1999).
%
%
\bibitem{pseudogap}
F.J. Ohkawa,
Phys. Rev. B {\bf 69},104502 (2004). 
%
%
%
\bibitem{ZhangRice}
F.C. Zhang and T.M. Rice,
Phys. Rev. B {\bf 37}, R3759 (1988). 
%
\bibitem{Gutzwiller} 
M.C. Gutzwiller, 
Phys. Rev. Lett. {\bf 10}, 159 (1963); Phys. Rev. A {\bf 134}, 293 (1963); 
 A {\bf137}, 1726 (1965).
%
\bibitem{Hubbard} 
J. Hubbard, 
Proc. Roy. Soc. London Ser. A {\bf 276}, 238 (1963); A {\bf 281}, 401 (1964).
%
\bibitem{slave}
F.J. Ohkawa,
J. Phys. Soc. Jpn. {\bf 58}, 4156 (1989).
%
\bibitem{Hirsch}
J.E. Hirsch,
Phys. Rev. Lett. {\bf 54}, 1317 (1985).
%
%
\bibitem{KL-theory1} 
F.J. Ohkawa,
J. Phys. Soc. Jpn. {\bf 56}, 2267 (1987).
%
\bibitem{KL-theory2} 
F.J. Ohkawa,
Phys. Rev. B. {\bf 59},  8930 (1990).
%
%
\bibitem{Mapping-1} 
F.J. Ohkawa, 
Phys. Rev. B {\bf 44}, 6812 (1991).
%
\bibitem{Mapping-2} 
F.J. Ohkawa, 
J. Phys. Soc. Jpn. {\bf 60}, 3218 (1991); {\bf 61}, 1615 (1992).
%
%
\bibitem{georges}
A. Georges and G. Kotliar,
Phys. Rev. B {\bf 45}, 6479 (1992).
%
\bibitem{metzner} 
W. Metzner and D. Vollhardt, 
Phys. Rev. Lett. {\bf 62}, 324 (1989).
%
\bibitem{gamma1} 
J.W. Loram, et al.,
Phys. Rev. Lett. {\bf 71}, 1740 (1993).
%
%
\bibitem{Davis}
K. McElroy, et al.,
Phys. Rev. Lett., {\bf 94}, 197005 (2005).
%
%
\bibitem{layer2}
R.A. Klemm, AIP Conf. Proc., n 273, 1993, p.~292.
%
%
\bibitem{layer3}
D. J\'{e}rome and H.J. Schultz,
Adv. Phys. {\bf 51}, 293 (2002).
%
\bibitem{ott}
H.R. Ott, et al.,
{\it Proc. of SPIE},
The International Society of Optical Engineering, Vol.~322, 
1999, p.~9.
%
\bibitem{chu}
C.W. Chu,
J. Superconductivity, {\bf 7}, 1 (1994).
\end{thebibliography}
\end{document}